\documentclass[runningheads]{cl2emult}

\def\p{\partial}
\def\s{\hat\sigma}
\def\r{\rho}
\def\a{\alpha}\def\b{\beta}
\def\C{{\cal C}}\def\Q{{\cal Q}}\def\N{{\cal N}}\def\A{{\cal A}}
\def\half{{\scriptstyle\frac{1}{2}}} 
\def\ihalf{{\scriptstyle\frac{i}{2}}}

\begin{document}

\title*{Emergence of Classicality: From Collapse \protect\newline 
Phenomenologies to Hybrid Dynamics\thanks{
To be published in the proceedings of the Bielefeld conference on
``Decoherence: Theoretical, Experimental, and Conceptual Problems",
edited by P. Blanchard, D. Giulini, E. Joos, C. Kiefer, and I.-O.
Stamatescu (Springer 1999).}
}

\toctitle{Emergence of Classicality: From Collapse 
\protect\newline Phenomenologies to Hybrid Dynamics}

\titlerunning{Emergence of Classicality}
 
\author{Lajos Di\'osi}

\authorrunning{Lajos Di\'osi}

\institute{
Institute for Advanced Study, Wallotstrasse 19, D-14193 Berlin, Germany
\and Research Institute for Particle and Nuclear Physics\\
	H-1525 Budapest 114, P.O.B. 49, Hungary}

\maketitle              

\begin{abstract}
In the past ten-fifteen years, stochastic models of continuous wave function
collapse were being proposed to describe the continuous emergence of 
classicality from quantum. We advocate that the hybrid dynamics of canonically
coupled quantum and classical systems is a more basic concept. Continuous 
collapse formalisms are obtained as special cases. To illustrate our claim we 
show how von Neumann collapse follows from hybrid dynamical equations.
\end{abstract}

\section{Introduction}
In 1978, Sherry and Sudarshan \cite{SheSud} wrote an 
unconventional paper on quantum measurements. "We treat the apparatus
as a classical system belonging to the macroworld. To describe
the quantum measurement process we must couple the classical
apparatus to the quantum system \dots we must first understand
how to engineer interactions between classical systems and
quantal systems" --- that is the idea. 
Indeed, the concept of dynamically interacting quantum and classical
systems could integrate all theoretical attempts to describe the emergence of 
classicality from quantum. Ideal quantum measurement, as well as spontaneous
emergence of classicality, would then be simple consequences of 
differential equations.

Yet, in the eighties research took a different path. It stuck,
firmly though not always explicitly, to the concept of quantum
measurement (collapse) \cite{Neu}. Typical proposals of dynamic or continuous
collapse mechanisms \cite{GRW,Bell,Dio89,Gis89,Pea89,DGHP}
fell into the class of continuous quantum measurements
\cite{Men76,Bar83,Dio88,CavMil,Bel,WisMil},
a more or less straightforward application of the standard
measurement theory, whereas the measuring apparatus is sometimes hypothetical 
or at least not to be identified. This was less straightforward ten years ago 
and it became more straightforward later. 

\section{From ideal to continuous collapses}
Von Neumann shows how his theory of ideal collapses applies to the indirect
measurement of any Hermitian observable \cite{Neu}.
To measure the position operator $\hat q$ of our quantized system $Q$ in its
quantum state $\psi(q)$, we let it to interact with the momentum $\hat p_\A$ 
of another quantum system $\A$ (ancilla) whose coordinate $\hat x_\A$ will be 
the pointer to show the measurement outcome $\bar q$. Accordingly, we assume a 
Gaussian wave function for $\A$, centered at $x_\A=0$ with precision $\Delta$:
\begin{equation}
\psi_\A(x_\A)=
[2\pi \Delta^2]^{-1/4}\exp\left[-\frac{x_\A^2}{4\Delta^2}\right].
\end{equation}
We assume that the system $Q$ and the ancilla $\A$ are uncorrelated initially,
the initial wave function factorizes as $\psi_\A(x_\A)\psi(q)$.
The system and the ancilla will interact only a very short time so that
the self--evolutions of $Q$ and $\A$ can be ignored during the measurement. 
We shall approximate the total Hamiltonian by
$
\delta(t)\hat q \hat p_\A~.
$
We can integrate the Schr\"odinger equation during the measurement. The 
factorized initial wave function transforms unitarily into the correlated one:
\begin{equation}
\psi_\A(x_\A)\psi(q)\rightarrow \psi_\A(x_\A -q)\psi(q).
\end{equation}
The pointer $\hat x_\A$ has taken over the value of the system coordinate 
$\hat q$. Let us read out the pointer's coordinate with a precision much
higher than $\Delta$ or any characteristic length of the system's
state $\psi(q)$. Hence we assume infinite precision formally. Then, 
according to von Neumann's collapse theory, the wave function of the ancilla
shrinks into a delta function $\delta(\bar q-x_\A)$, where $\bar q$ is the
measurement outcome, while the composite wave function (2) collapses
into the product of the ancilla's delta function and the systems's
new wave function:
\begin{equation}
\psi_\A(x_\A -q)\psi(q)\rightarrow 
\delta(x_\A-\bar q)\frac{\psi_\A(\bar q - q)\psi(q)}{\N(\bar q)}~.
\end{equation}
The factor $1/\N(\bar q)$ normalizes the system's new wave function:
\begin{equation}\label{N2}
\N^2(\bar q)=\int \vert\psi_\A(\bar q - q)\psi(q)\vert^2 dq~.
\end{equation}
Furthermore, the probability distribution of the outcome $\bar q$ is equal to 
the squared modulus of the overlap between the states respectively before (2) 
and after (3) the collapse:
\begin{eqnarray}
p(\bar q)&=&
\Big\vert \int \psi_\A^\star(x_\A -q)\psi^\star(q)\times\delta(x_\A-\bar q)
\frac{\psi_\A(\bar q - q)\psi(q)}{\N(\bar q)}dx_\A dq\Big\vert^2 =\nonumber\\
         &=&\N^2(\bar q)~.
\end{eqnarray}
In the second line we used Eq.~(4).
Now that the state (3) after the collapse factorizes again, we can summarize
the net effect of the above standard measurement on the system $\Q$, without 
any further reference to the ancilla $\A$. 
{\it The system's original wave function $\psi(q)$ has become multiplied by 
a Gaussian factor\/} [the ancilla's $\psi_\A(\bar q -q)$, in fact] {\it and 
then re--normalized:}
\begin{equation}
\psi(q)\rightarrow \N^{-1}(\bar q)
[2\pi \Delta^2]^{-1/4}\exp\left[-\frac{(\bar q-q)^2}{4\Delta^2}\right]
        \psi(q)
\end{equation}
{\it where the center $\bar q$ of the Gaussian} [the outcome of the standard
measurement] {\it is distributed according to the probability
distribution $p(\bar q)=\N^2(\bar q)$ being equal to the squared norm of the
unnormalized state after the collapse (6).}

Without referring to the above derivation from standard measurement theory,
this `hitting--process' had been {\it postulated} in the eighties 
\cite{GRW,Bell,Dio88} in order to build up {\it ad hoc} models of continuous 
emergence of classicality under various titles like continuous measurement, 
dynamical collapse, spontaneous collapse {\it e.t.c.\/}. Only few physicists 
\cite{CavMil} emphasized that the hitting--process was
formally derivable from standard measurement theory. On the contrary, many
thought that the process represented a modification of standard quantum 
theory. This belief made the proponents (including me, among others) 
enthusiastic since we sensed the flavor of a heuristically innovated and 
successful theory. The opponents drew the negative conclusion, warning that 
the modification of the standard theory was completely groundless \cite{Per}.  

Meanwhile the same mathematical equations of continuous measurement
were really obtained from standard quantum mechanics \cite{Bel,WisMil},
still in Markov approximation.
A few years later, however, it was possible to show that standard
quantum theory of atom+radiation, when described in proper basis, led
to exact stochastic equations for the atomic wave function \cite{Dio96}.
These equations, equivalent to the fully quantized theory on one hand,
turn out to reduce to the widely used phenomenological equations of
continuous (dynamical, spontaneous, whatever) collapse (measurement)
in the Markov limit. Indeed, the equations of continuous 
measurement {\it follow} from the hybrid representation
of standard quantum mechanics. 

\section{Hybrid dynamics and the ideal collapse}
The interaction between quantum and classical systems is called hybrid
dynamics \cite{Per}.
It was a long march from mean--field approximation \cite{Ros63} 
through its stochastic refinements \cite{HuMat,DioHal} and 
attempts at canonical coupling \cite{Ale,BouTra}
until the first mathematically consistent equations were written down 
\cite{Dio.Two,Dio.True}. 
We have finally obtained a general theory of hybrid dynamics \cite{Roy}.

Assuming a quantum system $\Q$ in state
$\hat\r_\Q$ and a classical canonical system $\C$ with phase space
distribution $\r_\C(x,p)$, we form the hybrid system $\Q\times \C$. If the
subsystems $\Q$ and $\C$ are uncorrelated then it is straightforward
to construct the hybrid state
\begin{equation}
\hat\r(x,p)=\hat\r_\Q\r_\C(x,p)
\end{equation}
for the composite system. In general, we represent the state of the hybrid 
system by a hybrid "density" $\hat\r(x,p)$ which is a phase space dependent
non--negative operator. Its trace is the phase space distribution 
$\r_\C(x,p)$ of $\C$ while its phase space integral yields the density
operator $\hat\r_\Q$ of $\Q$. When $\hat\r(x,p)$ is not factorable the 
unconditional quantum state $\hat\r_\Q$ must be distinguished from the 
conditional quantum states:
\begin{equation}
\hat\r_{xp}=\frac{\hat\r(x,p)}{\r_\C(x,p)}
\end{equation}
depending on the classical coordinates $x,p$ as conditions. 
The Hamiltonian of the hybrid system takes this form:
\begin{equation}
\hat H(x,p)=\hat H_\Q + H_\C(x,p) +\hat H_{INT}(x,p)
\end{equation}
where, obviously, the interaction term is a phase space dependent Hermitian 
operator. One can construct the following canonical hybrid equation of 
motion \cite{Ale} for the hybrid state $\hat\r(x,p)$:
\begin{equation}
\p_t\hat\r=-i[\hat H,\hat\r]+\half\{\hat H,\hat\r\}_P
                            -\half\{\hat\r,\hat H\}_P
\end{equation} 
which is the naive combination of the Dirac $[~,~]$ and the Poisson
$\{~,~\}_P$ brackets. Unfortunately, this equation does not preserve
the positivity of $\hat\r(x,p)$. So, the naive construction (10) does not
work. In fact, the hybrid dynamics cannot be a true reversible
dynamics. We have to make a little compromise. This I found first
for the special case when $\hat H_{INT}$ is linear in $x$ and $p$ 
\cite{Dio.Two}. One applies the following Gaussian coarse graining, 
over Planck cells, to the hybrid state:
\begin{equation}
\hat\r(x,p)\rightarrow\int\exp[-(\xi^2 + \eta^2)]\hat\r(x+\xi,p+\eta)
\frac{d\xi d\eta}{2\pi}~.
\end{equation}
Applying this coarse--graining on both sides of the naive equation (10),
one obtains two new terms:
\begin{equation}
\p_t\hat\r=-i[\hat H,\hat\r]+\half\{\hat H,\hat\r\}_P
                            -\half\{\hat\r,\hat H\}_P
                            -\ihalf[\p_x\hat H,\p_x\hat\r]
                            -\ihalf[\p_p\hat H,\p_p\hat\r]~.
\end{equation}
And this equation, as can be shown, preserves the positivity of the 
coarse--grained hybrid state. Of course, we cannot choose an arbitrary 
hybrid state as initial state. E.g., sharp values of $x$ and $p$, or wild
fluctuations within single Planck cells are forbidden. The rigorous 
constraints for $\hat\r(x,p)$ are given elsewhere \cite{Dio96,Roy}.

By taking the trace of Eq.~(12), one can show that the evolution of the
classical states is a flow:
\begin{equation}
\p_t x= \left\langle \p_p\hat H(x,p) \right\rangle_{xp}~,~~
\p_t p=-\left\langle \p_x\hat H(x,p) \right\rangle_{xp}~,
\end{equation}
where $\langle\dots\rangle_{xp}$ stands for the expectation values 
$tr(\dots\hat\r_{xp})$ in the current conditional
quantum state (8). This flow generalizes the naive mean--field equations
where $\hat\r_{xp}\equiv\hat\r_\Q$ and quantum fluctuations are ignored
in the back--reaction of the quantum system $\Q$ on the classical $\C$.

But there are other earlier concepts which are recovered by hybrid dynamics.
Quantum Brownian motion is one.
The exact non--Markov stochastic Schr\"odinger--equation \cite{DioStr,SDG}
of the Caldeira--Leggett--type open systems (which include, e.g., the 
atom+radiation systems) follows automatically from the corresponding hybrid
equations (12) \cite{Dio96,Roy}. This means that, in particular, the 
phenomenological Ito--Schr\"odinger--equations of continuous (dynamical) 
collapse (measurement) follow from the hybrid equations in the Markov limit.

Finally I demonstrate the "presence" of collapse mechanism in the hybrid
dynamics. To this end, I show that the hybrid dynamical equations (12)
describe the Stern--Gerlach measurement, including the collapse of the spin's
state and the corresponding motion of the classical pointer. 
Our quantum system $\Q$ is the electron's spin and initially it is in the 
superposition
\begin{equation}
\vert in\rangle=c_+\vert+\rangle+c_-\vert-\rangle
	       =\sum_{\a=\pm1}c_\a\vert\a\rangle
\end{equation}
of the two eigenstates $\vert\pm\rangle$ of $\s_3$. Our classical system 
$\C$ is the pointer. Let it be a harmonic oscillator with Hamiltonian 
$\half(x^2+p^2)$, shortly but strongly coupled to the measured spin 
component $\s_3$ by the interaction $\hat H_{INT}=g\delta(t)p\s_3$,
where $\Delta=1/g$ will be the precision of the measurement and we assume
$\Delta\ll1$. Since we are interested in the states just before and, 
respectively, after the measurement, only the interaction Hamiltonian is 
relevant and the hybrid equation (12) will take this form:
\begin{equation} 
\p_t\hat\r=
-ig\delta(t)p[\s_3,\hat\r]-\half g\delta(t)[\s_3,\p_x\hat\r]_{+}
-\ihalf g\delta(t)[\s_3,\p_p\hat\r]~.
\end{equation}
As it follows from this dynamics, $\bar\sigma_3=x/g=x\Delta$ will play the 
role of the pointer variable to indicate the value of the spin operator 
$\s_3$ after the measurement. We assume the following factorized initial 
state for the hybrid system:
\begin{equation}
\vert in\rangle\langle in\vert\frac{\exp[-\half(x^2+p^2)]}{2\pi}\equiv
\vert in\rangle\langle in\vert\r_\C(x,p;in)~,
\end{equation}
where $\r_C(x,p;in)$ corresponds to the pointer's initial position $x=0\pm1$,
i.e. to $\bar\sigma_3=0\pm\Delta$. The evolution (14) acts on matrix elements 
of the initial state (16) as follows:
\begin{eqnarray}
&&\exp\Bigl(-igp[\s_3,~.]-\half g[\s_3\p_x,~.]_{+}
           -\ihalf g[\s_3\p_p,~.]\Bigr)
\vert\a\rangle\langle\b\vert\r_\C(x,p;in)=\nonumber\\
&=&\vert\a\rangle\langle\b\vert
\exp\left(-\frac{(\a-\b)^2}{4}g^2-i(\a-\b)gp\right)\nonumber\\
&&\r_\C\left(x-\frac{\a+\b}{2}g,p+\frac{\a-\b}{2i}g;in\right)~.
\end{eqnarray}
We see that the off--diagonal terms are heavily damped, so the initial state
(16) transforms into a diagonal final state:
\begin{eqnarray}
\sum_{\a=\pm1} \vert c_\a\vert^2\vert\a\rangle\langle\a\vert
	                                  \r_\C(x-\a g,p;in)\nonumber\\
=\vert c_+\vert^2 \vert +\rangle\langle+\vert\r_\C(x-g,p;in) 
+\vert c_-\vert^2 \vert -\rangle\langle-\vert\r_\C(x+g,p;in)~. 
\end{eqnarray}
This result clearly shows that the pointer's coordinate shifts either to the 
right $(x=g\pm1)$ with probability $\vert c_+\vert^2$ and then the spin's
state is $\vert+\rangle$, or it moves to the left $(x=-g\pm1)$ in the 
complementary cases:
\begin{equation}
\left(\vert out\rangle, \bar\sigma_3\right) = 
\cases{\left(\vert +\rangle,~\bar\sigma_3=+1\pm\Delta\right) &
                               ~~~with probability $\vert c_+\vert^2$\cr
       \left(\vert -\rangle,~\bar\sigma_3=-1\pm\Delta\right) &
                               ~~~with probability $\vert c_-\vert^2$}
\end{equation}	   
where $\Delta\ll1$. This scheme of the final quantum {\it and} classical 
pointer states is, regarding to the initial state (16) with the superposed 
spin (14), identical to the result of the corresponding ideal \cite{Neu}
Stern--Gerlach quantum measurement.

\section{Summary}
As I argued in Sec.~2, all phenomenological stochastic 
Schr\"odinger equations, however sophisticated they are, remain in the
framework of standard quantum mechanics (whose part is the von Neumann 
measurement theory, too). This shall of course question
part of the criticism that these proposals are groundless modifications of
quantum mechanics since they are not modifications after all. Rather they 
are indicating the natural presence of continuous collapse mechanisms 
within standard quantum theory. 

In Sec.~3 I illustrated that the concept of canonically interacting 
classical and quantum systems automatically implies
the emergence of classicality in a way which is definitely more
general than the concept of collapse (measurement). Ideal collapses, 
continuous (Markov or non--Markov) collapses follow from the hybrid
dynamics. The paradigmatic (and controversial) mean--field approach can 
naturally be identified and improved within the hybrid dynamics. 

Like all continuous collapse models, also hybrid dynamics is equivalent
mathematically with a certain enlarged unitary dynamics. Hybrid dynamics is
a powerful unified framework to describe the variety how classicality 
'appears' \cite{Zeh} from quantum, yet this new phenomenology is in itself
unlikely to innovate our knowledge about the foundations.
We are being captured in the old castle of standard quantum
mechanics. Sometimes we think that we have walked into a new wing.
It belongs to the old one, however. 

\vskip 1truecm
\noindent {\bf Acknowledgments}

\noindent I thank the conference organizers for the invitation. This work was 
partly supported by the Hungarian Scientific Research Fund under 
Grant No. T016047.

\clearpage
\addcontentsline{toc}{section}{Index}
\flushbottom

\end{document}